\documentclass[12pt,a4paper]{iopart}

\newcommand{\mybe}{$^9\mathrm{Be}^+\,$}		
\usepackage[utf8]{inputenc}
\usepackage{blindtext}
\usepackage{hyperref}
\usepackage{nicefrac}
\usepackage{graphicx}
\usepackage{siunitx}
\usepackage{graphicx}
\usepackage{braket}
\usepackage{subfig}
\begin{document}
\title{Cryogenic $^9\text{Be}^+$ Penning trap for precision measurements with (anti-)protons}
\author{M. Niemann$^{1,2}$, T. Meiners$^{1,2}$, J. Mielke$^{1,2}$, M. J. Borchert$^{1,3}$, J. M. Cornejo$^{1,2}$, S. Ulmer$^3$ and C. Ospelkaus$^{1,2,4}$}
\address{$^1$ Institut f\"ur Quantenoptik, Leibniz Universit\"at Hannover, Welfengarten 1, 30167 Hannover, Germany}
\address{$^2$ Laboratorium f\"ur Nano- und Quantenengineering, Leibniz Universit\"at Hannover, Schneiderberg 39, 30167 Hannover}
\address{$^3$ Ulmer Fundamental Symmetries Laboratory, RIKEN, Wako, Saitama 351-0198, Japan}
\address{$^4$ Physikalisch-Technische Bundesanstalt Braunschweig, Bundesallee 100, 38116 Braunschweig, Germany}
\ead{christian.ospelkaus@iqo.uni-hannover.de}

\date{\today}

\begin{abstract}
Cooling and detection schemes using laser cooling and methods of quantum logic can contribute to high precision CPT symmetry tests in the baryonic sector.
This work introduces an experiment to sympathetically cool protons and antiprotons using the Coulomb interaction with a \mybe ion trapped in a nearby but separate potential well.
We have designed and set up an apparatus to show such coupling between two identical ions for the first time in a Penning trap.
In this paper, we present evidence for successful loading and Doppler cooling of clouds and single ions are presented.
Our coupling scheme has applications in a range of high-precision measurements in Penning traps and has the potential to substantially improve motional control in these experiments.
\end{abstract}

\maketitle

\section{Introduction}
Precise comparisons of the fundamental properties of matter/antimatter conjugates on the single particle level provide sensitive tests of CPT symmetry and the associated Lorentz symmetry in the Standard Model of Particle Physics \cite{luders_proof_1957,greenberg_$cpt$_2002}.
Important examples for this are comparisons of charge-to-mass ratios and magnetic moments of antiprotons and protons in Penning traps, where much experimental progress has been made recently \cite{smorra_350-fold_2018,higuchi_progress_2018,schneider_double-trap_2017,ulmer_high-precision_2015}.
However, especially in case of the most precise magnetic moment measurements \cite{schneider_double-trap_2017}, which rely on the double Penning trap technique \cite{haffner_double_2003}, the efforts are impeded by the absence of fast and deterministic preparation schemes for the motional state of the particle.
Standard laser-based manipulation and cooling techniques well established in atomic physics cannot be applied directly to a single proton or antiproton, as the nuclear spin is their only internal degree of freedom.
Therefore, current state of the art experiments apply sub-thermal resistive cooling methods, where the particle is coupled to cryogenic resistors \cite{borchert_measurement_2019}.
These time-consuming procedures are ultimately required to prepare particles with motional temperatures at the level of $T\leq 0.15\,$K \cite{smorra_observation_2017}, which is essential for the high-fidelity spin state detection carried out in these experiments.
In our currently most precise experiments, the sub-thermal cooling process constitutes the biggest time contribution in the measurement scheme.
Therefore, such experiments would considerably profit from the development of laser-cooling schemes, which deterministically provide particles at mode-temperatures below the single-spin-state detection threshold within much reduced time.
This is the inspiration and motivation to implement methods to sympathetically cool protons and antiprotons by coupling them to a co-trapped `logic' ion that can be manipulated.
Laser-cooling of the `logic'-ion and subsequent mode energy transfer will provide deterministically cold protons/antiprotons on time-scales of seconds.
This will outperform the currently used methods due to considerably improved sampling statistics and the lower particle energies achieved, which will also further reduce systematic uncertainties.
In future experiments, quantum logic inspired manipulation techniques could also be applied for state readout.
Moreover, any techniques developed to mitigate this could well be extended to other charged particles that are difficult to manipulate directly, for example to highly charged ions for ultra-stringent tests of bound state quantum electrodynamics.
Different coupling schemes have been proposed \cite{heinzen_quantum-limited_1990,wineland_experimental_1998} and are being worked on \cite{brown_coupled_2011,harlander_trapped-ion_2011,niemann_cpt_2013,smorra_base_2015,cornejo_optimized_2016,bohman_sympathetic_2018}.
This article describes the necessary steps to apply sympathetic cooling of protons and antiprotons via direct Coulomb coupling in an advanced Penning trap system.
Towards this end, we demonstrate operation of a cryogenic \mybe Penning trap which supplies the laser cooled ions for sympathetic cooling and quantum logic spectroscopy.

\section{Coupling concept}

\begin{figure}
	\begin{center}
	\includegraphics[width=.47\columnwidth]{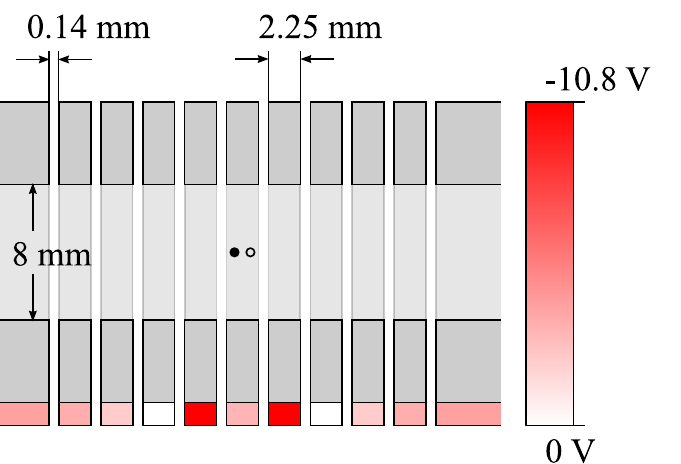}
	\hfill
	\includegraphics[width=.47\columnwidth]{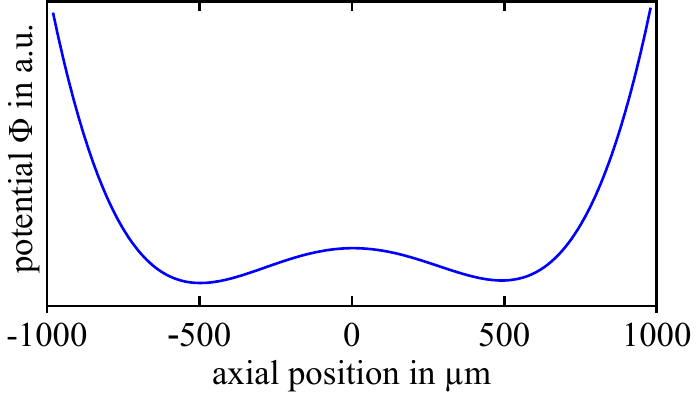}
	\end{center}
	\caption[Figure 1]{(a) Cut view of the cylindrical Penning trap for coupling of axial motional degrees of freedom. The applied trap voltages are color-coded below. (b) Potential shape of the double-well potential along the $z$-axis for motional coupling of two identical ions. This potential can be generated with the trap geometry shown in (a) and yields an exchange time $\tau_{\mathrm{ex}}=\,$\SI{64}{\milli\second} at an axial trap frequency $\omega_z=2\pi\cdot$\SI{100}{\kilo\hertz}.}
	\label{fig1}
\end{figure}

For charged particles, the Coulomb interaction provides the strongest coupling between particles and makes a natural choice for sympathetic cooling and state transfer.
Proposals to exploit this idea for sympathetic cooling and state detection have been made using different approaches, namely trapping the particles in a common potential well \cite{larson_sympathetic_1986}, a shared-electrode-approach \cite{heinzen_quantum-limited_1990} and a free-space approach where the ions are located in separate potential wells \cite{wineland_experimental_1998}.
As particles of opposite charge cannot be trapped in a common electrostatic potential well, we pursue the direct coupling approach in order to be able to address negatively charged particles as well.
This approach has been successfully demonstrated in radio-frequency surface traps for particles of equal charge-to-mass ratio \cite{brown_coupled_2011,harlander_trapped-ion_2011}.

In a Penning trap, charged particles are trapped by superimposing a strong magnetic field along the $z$-direction with a static electric quadrupole field confining the motion along that axis.
The resulting motion can be described by three modes: A harmonic oscillation with frequency $\nu_z$ along the $z$-axis called the axial mode and two radial modes called magnetron and modified cyclotron modes with frequencies $\nu_-$ and $\nu_+$ \cite{brown_geonium_1986}.

For two charged particles with charges $q_{1,2}$ trapped in separate harmonic potential wells along the $z$-axis with their equilibrium positions separated by $d_0$ and displacements $z_{1,2}$ from their respective equilibrium positions, the Coulomb interaction is described by the potential
\begin{equation}
U(z_1,z_2)= \frac{1}{4\pi\varepsilon_0} \frac{q_1 q_2}{d_0-\left(z_1-z_2\right)},
\end{equation}
where $\varepsilon_0$ is the vacuum permittivity.
Following the lines of \cite{brown_coupled_2011} and using the standard ladder operators $a_i$ and $a_i^\dagger$, the first order interaction can be expressed as
\begin{equation}
\frac{q_1 q_2}{2 \pi \varepsilon_0 {d_0}^3}\left( z_1 z_2 \right) \approx \hbar\Omega_{\mathrm{ex}}\left( a_1 a_2^\dagger + a_1^\dagger a_2 \right) ,
\end{equation}
where the last step assumes the angular frequencies $\omega_1$ and $\omega_2$ being close to each other.
Here,
\begin{equation}
\Omega_{\mathrm{ex}}=\frac{q_1 q_2}{4 \pi \varepsilon_0 {d_0}^3 \sqrt{m_1 m_2} \sqrt{\omega_1 \omega_2}}
\label{eqn:coupling_rate}
\end{equation}
is the coupling rate, $m_1$, $m_2$ denote the respective masses of the particles and $\hbar$ is the reduced Planck constant. For resonant (identical) trap frequencies $\omega_1=\omega_2=\omega_0$, the oscillators 1 and 2 swap energies with a period of $2\tau_{\mathrm{ex}}$, and exchange time $\tau_{\mathrm{ex}}=\pi/(2\Omega_{\mathrm{ex}})$. \\
For given particles, the coupling rate \eref{eqn:coupling_rate} scales strongly with the interparticle distance $d_0$, and to a lesser extent with the axial trap frequencies $\omega_{1,2}$. It would be favorable to combine low motional frequencies with a small distance between potential wells, but as decreasing the length scale of a two-well potential increases the gradient and curvature of the potential, the motional frequencies will scale unfavorably. Due to the different power law scaling of $\Omega_{\mathrm{ex}}$ with $d_0$ and $\omega_0$, a system with small length scales and a tighter confinement would be useful, but will ultimately demands for a miniaturised Penning trap that is currently under development. The version of the Penning trap which is installed for the first demonstration experiments presented in this manuscript is shown in Figure \ref{fig1}.
In order to achieve a stable coupling rate during the energy exchange periods (tens of milliseconds), axial frequency fluctuations should be small compared to the coupling rate (several or several tens of Hz).
The exchange times are fast compared to the expected ultra-low heating rates \cite{borchert_measurement_2019} in our cryogenic Penning trap system.

\section{Experimental setup}
The design of the experimental apparatus follows constraints defined by the apparatus of the BASE experiment, located at the antiproton decelerator of CERN \cite{smorra_base_2015}. This ensures mechanical compatibility of our system with the facilities at CERN to allow applying the demonstrated techniques on antiprotons in the future.\\
\subsection*{Superconducting magnet}
The magnetic field for trapping ions is supplied by a high homogeneity wet superconducting magnet supplied by \textit{Oxford Instruments} identical to the one described in \cite{smorra_base_2015}, but operated at a higher magnetic field of $B_0=5\,\mathrm{T}$. The experiment is placed in the horizontal room temperature bore.
The magnet's characterization results were in line with the data in \cite{smorra_base_2015} with a spatial homogeneity $\Delta B/B<0.3\cdot10^{-6}$ in a $1\,\mathrm{cm}^3$ volume, less than $4\cdot10^{-5}$ over the trap stacks' extent (about \SI{20}{\centi\meter}) and a temporal stability of $(\Delta B/B)(1/\Delta t)<4\cdot10^{-9}/\mathrm{h}$.

\subsection*{Cryo-mechanical setup}

\begin{figure}
	\centering
	\includegraphics[width=\columnwidth]{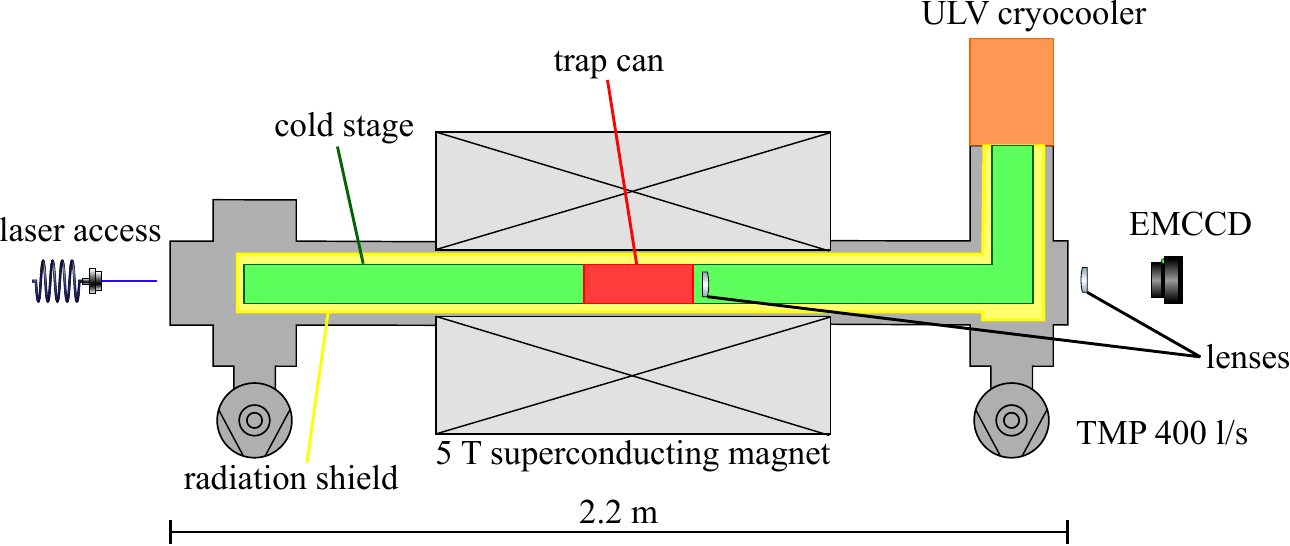}
	\caption[Figure 2]{Simplified schematic overview of the experimental layout. The trap can (red) is placed in the homogeneous region of the superconducting magnet. It is supported and cooled by a cryomechanical structure made of OFHC and titanium (green). This structure is supported and shielded by a radiation shield mostly made from aluminium (yellow). The vacuum chamber is continuously pumped by two turbomolecular pumps. Laser beams are introduced from the left side, whereas the imaging system and wiring is on the right side of the picture.}
	\label{fig3b}
\end{figure}
An overview of the vacuum system is shown in Figure \ref{fig3b}.
A horizontal bore magnet was chosen in order to be mechanically compatible with the BASE CERN setup.

The trap and wiring system form a separate cryogenic system independent from the temperature of the superconducting magnet.
This allows for maintenance and upgrades on the experiment while maintaining the carefully shimmed magnetic field.
All materials used in the magnet have an extremely low magnetic permeability to keep the field as homogeneous as possible.
The main cold stage is cooled by the second stage of a two-stage Gifford-McMahon cryocooler equipped with a helium gas heat exchanger for vibration isolation (\textit{ColdEdge Techologies SRDK-415 ULV UHV}) between the cold head and experiment.
Our group measured an identical cooler to have vibration amplitudes below \SI{25}{\nano\meter} at the cold plate.
Homemade flexible braids made from oxygen free high conductivity copper (OFHC) further reduce vibrations transmitted to the cold temperature stages and allow for thermal contraction in the apparatus.
An aluminium radiation shield is thermally anchored to the first cooling stage of the cryocooler.
It is mechanically fixed to the magnet using a titanium structure and disc-shaped G10CR (a glass fiber filled epoxy for low temperature use) spacers with a maze structure to minimize heat leakage.
The calculated conductive heat load is about two percent of the cryocooler's cooling power at \SI{4}{\kelvin}, rendering it negligible.
The cryogenic trap system and wiring is mounted to the radiation shield using a similar structure.
To improve thermal contact, a vacuum annealed copper rod of $99.999\%$ purity is used for thermal contact between the cooler interface and the trap system.
Radiative heat transfer is suppressed using high-reflectivity aluminium foil and multi-layer insulation blankets.
Convective heat transfer is minimized by pumping the vacuum system to less than $10^{-7}\,$mbar with two turbomolecular pumps before cooling down.
With this system, we achieve temperatures of around \SI{5}{\kelvin} near the trap.\\
Laser beam access is realized using custom in-vacuum mirrors to direct the beam through and out of the vacuum system.
All laser beam adjustment is done outside of the vacuum chamber on a dedicated platform.

\subsection*{Penning trap system}

\begin{figure} 
	\centering
	\includegraphics[width=\columnwidth]{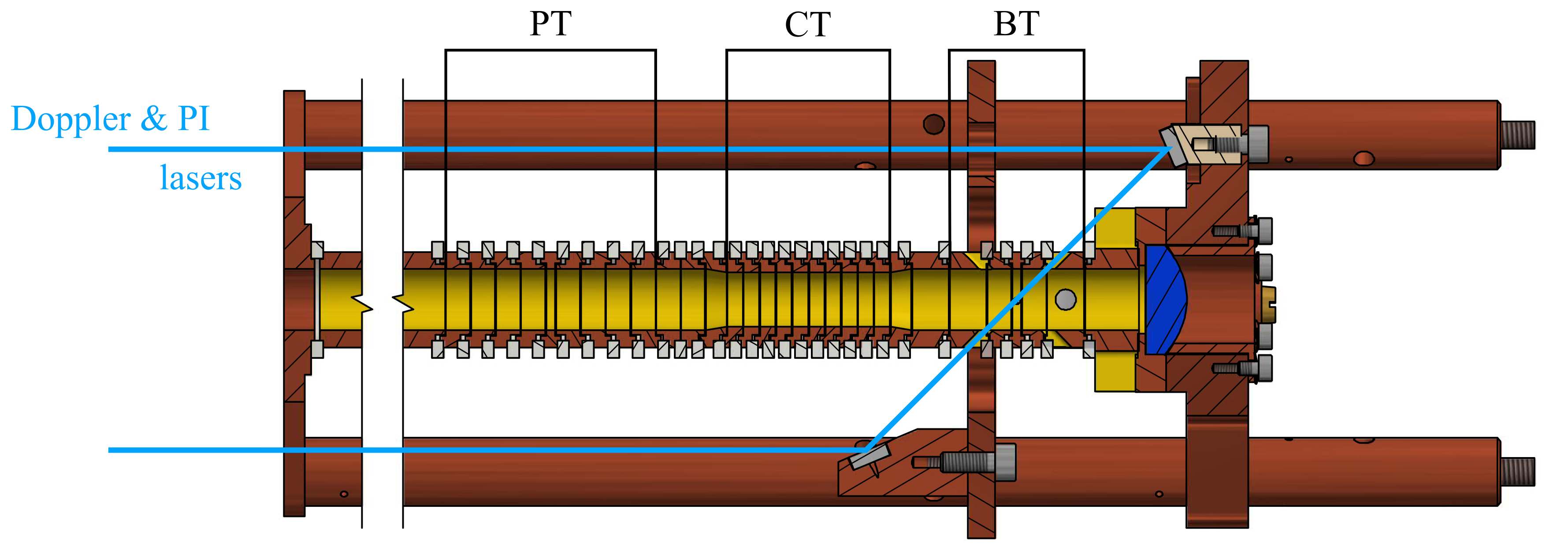}\\
	\includegraphics[width=\columnwidth]{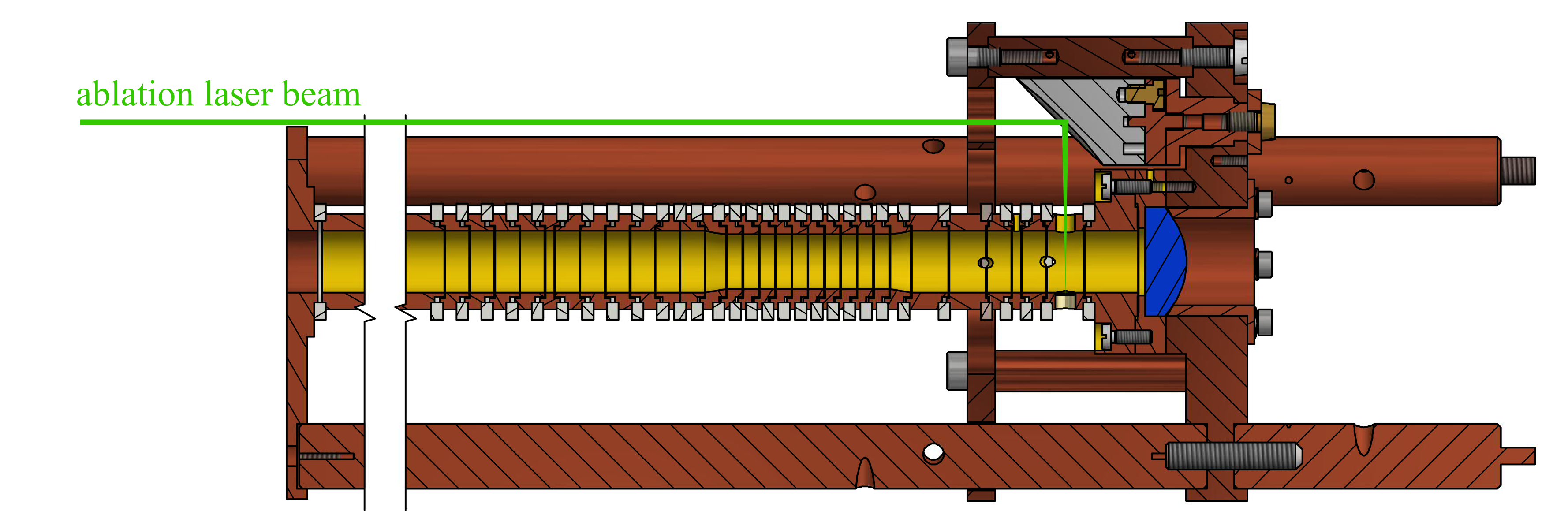}
	\caption[Figure 3]{Trap stack and laser layout. PT: precision trap. CT: coupling trap. BT: beryllium trap. The inner trap diameter is \SI{8}{\milli\meter} for the CT and \SI{9}{\milli\meter} for the other traps. (a) shows the UV laser beam path that crosses the trap axis at a $45^\circ$ angle and is directed out of the apparatus for beam position monitoring. (b) shows the ablation laser beam path which runs in a plane perpendicular to the plane in (a). The beam is introduced collimated and is focused by an off-axis parabolic mirror (grey) onto a beryllium disc to produce a plume of atoms.} 
	\label{fig4}
\end{figure}

The current trap system (see Figure \ref{fig4}) consists of three cylindrical Penning traps with open endcaps located in a cylindrical OFHC `trap can'.
In order to reach the extreme high vacuum (XHV) regime necessary to prevent antiproton annihilation, the trap can can be hermetically sealed to form its own cryopumped vacuum system \cite{sellner_improved_2017}.
The trap stack is comprised of a 'beryllium trap', a coupling trap and a precision trap.
Extra space for additional traps, such as a proton source trap, is currently reserved with a dummy electrode.
All trap electrodes are precision-machined from OFHC copper and gold-plated to prevent oxidation and minimize surface charge effects.
Electrodes are separated by sapphire rings that also ensure concentric alignment of the assembly.
\\
The beryllium trap (BT) is a compensated and orthogonal five-electrode Penning trap with laser access and loading capabilities.
Single pulses from a \SI{532}{\nano\meter} laser are tightly focused onto a solid beryllium target embedded in an endcap electrode using an in-vacuum off-axis parabolic (OAP) mirror to produce beryllium atoms and ions.
Laser beams for photoionization, cooling and repumping are introduced using custom in-vacuum mirrors and at an angle of $45^\circ$ with respect to the trap axis to allow coupling to all motional modes.
The laser beam is guided out of the vacuum chamber along a parallel path to facilitate adjustment of the beam and allow for beam position monitoring.
A custom-made aspheric lens (numerical aperture NA=0.25) collects fluorescence light, which is focused and directed onto either an EMCCD camera (\textit{Andor iXon 885}) or a photomultiplier tube (\textit{Hamamatsu H8259-01}) for state detection using a lens outside of the vacuum system.\\
The coupling trap (CT) is a stack of ten equally sized electrodes that allows for applying a double-well potential suitable for coupling the axial motional modes of two ions.
For the first demonstration of the direct motional coupling of two beryllium ions in a double-well potential, we will use a trap geometry with an inner diameter of \SI{8}{\milli\meter}, an inter-ion distance of $d_0=\,$\SI{1}{\milli\meter} and $\omega_0/(2\pi)=\,$\SI{100}{\kilo\hertz}, resulting in an exchange time $\tau_{\mathrm{ex}}\approx\,$\SI{64}{\milli\second} and a coupling rate $\Omega_{\mathrm{ex}}\approx\,$\SI{25}{\per\second}.\\
The precision trap (PT) is a compensated, orthogonal five-electrode trap \cite{smorra_base_2015} that can be used for high-precision radio-frequency spectroscopy of protons.
All endcap electrodes are segmented into multiple segments to allow for adiabatic transport between traps.
Traps are connected with transport electrodes and the segments between the coupling electrodes and the other traps are tapered.

Trap voltages are supplied by a homebuilt voltage source \cite{bowler_arbitrary_2013} and amplifiers (\textit{APEX PA98}). They are connected to the electrodes using low-pass filters on all three temperature stages to reduce noise and ensure thermalization of the lines.
The DC voltage source used is capable of a \SI{50}{\mega\hertz} update rate and limited by low-pass filters with a total cut-off frequency of \SI{3.1}{\kilo\hertz} outside the vacuum chamber and on the cryogenic stages.
Additional coaxial lines allow for axial excitation using an endcap electrode and radial excitation using a segmented correction electrode.
The system is prepared to be upgraded with image current detection systems consisting of tank circuits and ultra-low noise amplifiers to enable resistive cooling and motional frequency measurements on protons \cite{nagahama_highly_2016}.

\subsection*{Laser systems}
\begin{figure}
	\centering
	\includegraphics[width=\columnwidth]{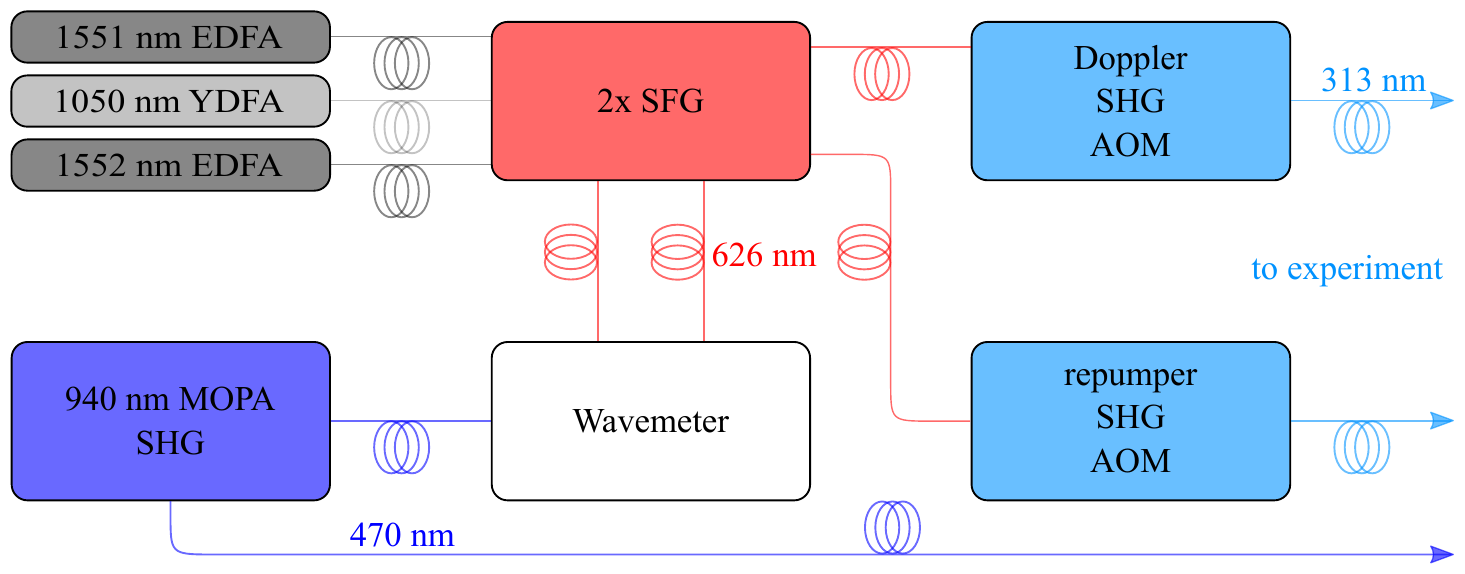}
	\caption[Figure 4]{Schematic layout of the laser systems. The fiber amplifiers are located in a rack, while each of the other rectangular boxes in this picture represents an optical module based on aluminium breadboards of \SI{60}{\centi\meter} by \SI{60}{\centi\meter} size. These modules are stackable and interconnected by optical fibers. Acousto-optic modulators (AOMs) are used for intensity stabilization.}
	\label{fig5}
\end{figure}
The laser setup is shown in \fref{fig5}.
It is comprised of an ablation laser for ablation of atoms and ions, a photoionization (PI) laser for resonantly enhanced photoionization in neutral $^9$Be, a Doppler cooling and a repumping laser.\\
The PI laser adresses the $\left|^1\mathrm{S}_0,\,m_J=0\right\rangle\rightarrow\,\left|^1\mathrm{P}_1,\;m_J=0\right\rangle$ transition near \SI{235}{\nano\meter}.
The Doppler and repumper lasers are tuned to the $\ket{^2\mathrm{S}_{\nicefrac{1}{2}},\,m_J=\nicefrac{1}{2},\,m_I=\nicefrac{3}{2}}\rightarrow\;\ket{^2\mathrm{P}_{\nicefrac{3}{2}},\,m_J=\nicefrac{3}{2},\,m_I=\nicefrac{3}{2}}$ and $\ket{^2\mathrm{S}_{\nicefrac{1}{2}},\,m_J=-\nicefrac{1}{2},\,m_I=\nicefrac{3}{2}}\rightarrow\;\ket{^2\mathrm{P}_{\nicefrac{3}{2}},\,m_J=\nicefrac{1}{2},\,m_I=\nicefrac{3}{2}}$ transitions, both close to \SI{313}{\nano\meter}.
The ablation laser is a commercial ns-pulsed, frequency doubled Nd:YAG laser near \SI{532}{\nano\meter} (\textit{Continuum Minilite}).
The photoionization laser is a \SI{940}{\nano\meter} master oscillator power amplfifier (MOPA) system (\textit{TOPTICA TA pro}) that is frequency-quadrupled in two doubling stages following \cite{wilson_750-mw_2011,lo_all-solid-state_2014,hannig_highly_2018}.
The Doppler and repumping laser beams are generated by sum frequency generation (SFG) and subsequent second harmonic generation (SHG) using the output of fiber laser systems with an ytterbium doped fiber amplifier (YDFA) close to \SI{1050}{\nano\meter} and two erbium doped fiber amplifiers (EDFA) close to \SI{1550}{\nano\meter} (\textit{NKT Photonics Koheras Adjustik \& Boostik}), also following the general layout of \cite{wilson_750-mw_2011,lo_all-solid-state_2014}, using the cavities described in \cite{hannig_highly_2018}.
The \SI{313}{\nano\meter} laser light and the intermediate \SI{470}{\nano\meter} light for the \SI{235}{\nano\meter} source are guided from an optical table to a laser platform near the experiment using polarization-maintaining photonic crystal fibers (\textit{NKT LMA-PM-10}, prepared as described in \cite{colombe_single-mode_2014,marciniak_towards_2017}).
The final frequency doubling stage \SI{470}{\nano\meter}$\,\rightarrow\,$\SI{235}{\nano\meter}, the ablation laser source and beam shaping optics are located on that same platform.
See Figure \ref{fig5} for a schematic layout of the laser setup.
All laser sources but the ablation laser are frequency stabilised by a wavelength meter (\textit{High Finesse WSU-2}) that is referenced to a calibrated HeNe laser source.
We measured the accuracy of the wavemeter stabilisation to be better than \SI{2}{\mega\hertz} over a six-hour period by beating a \SI{626}{\nano\meter} laser with another one that was locked on an iodine reference via frequency modulation (FM) spectroscopy.

\section{Results}
\subsection{Trap loading}

\begin{figure} 
	\centering
	\subfloat[]{\includegraphics[width=.3\columnwidth, trim= 107 212 2332 156, clip=true]{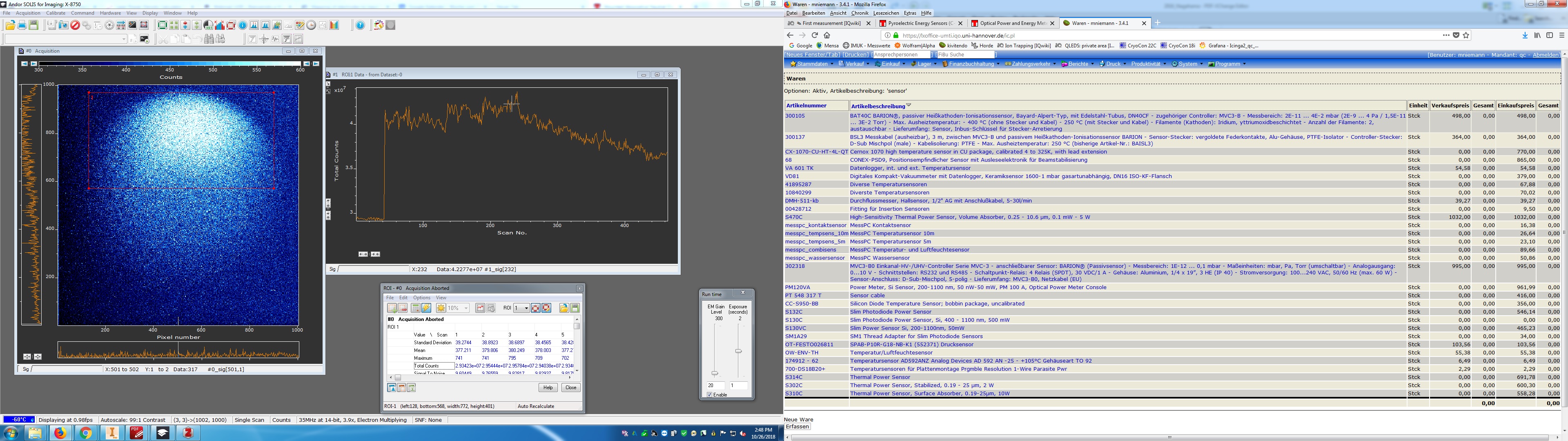}} 
	\hspace{.01\columnwidth}
	\subfloat[]{\includegraphics[width=.3\columnwidth, trim= 83 206 2356 162, clip=true]{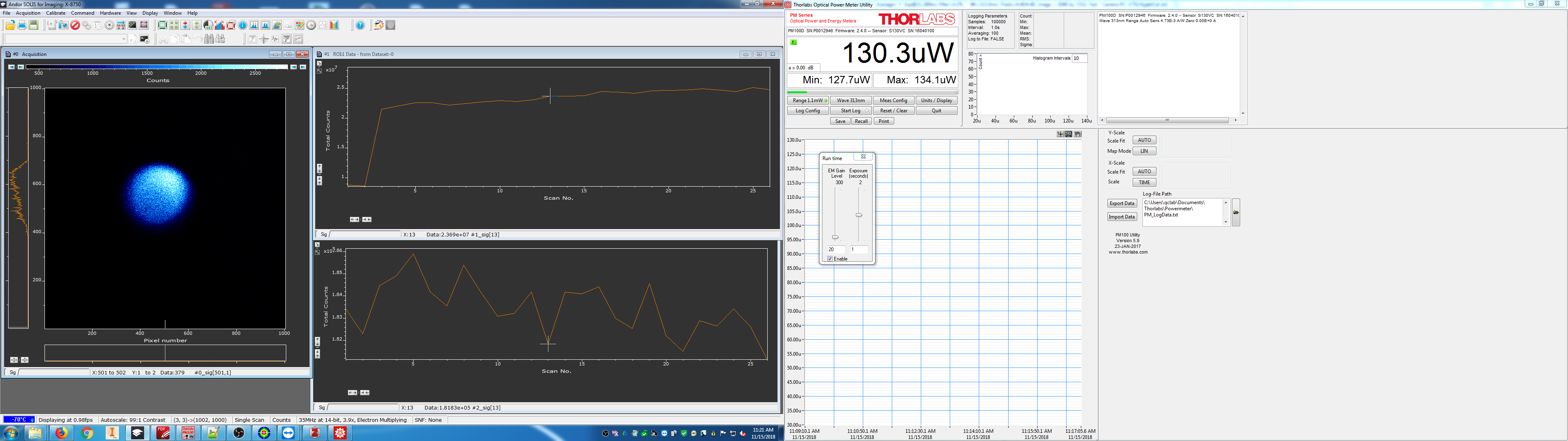}} 
	\hspace{.01\columnwidth}
	\subfloat[]{\includegraphics[width=.3\columnwidth, trim= 83 206 2356 162, clip=true]{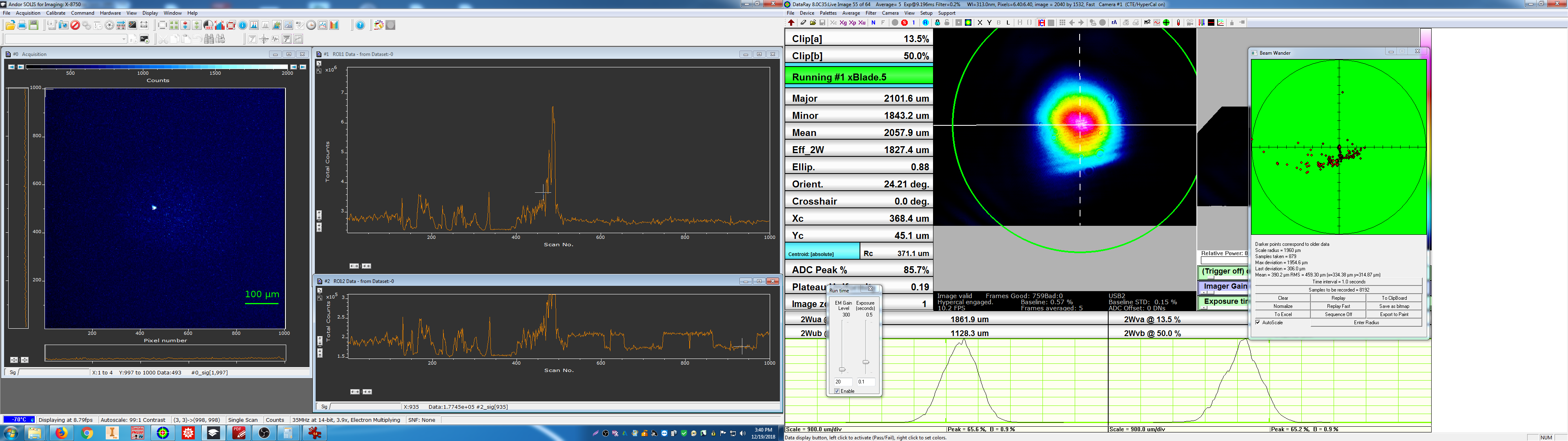}} 
	\caption[Figure 5]{Fluorescence images of \mybe clouds illuminated with a near-resonant Doppler laser. (a) First trapped cloud with unoptimised imaging system. (b) Large cloud with optimised imaging system. The imaged cloud is larger than can be seen here due to clipping on apertures along the imaging path. The inhomogeneity of the brightness is due to laser beam alignment. (c) A single ion. All pictures use the same length scale, but different intensity scaling.} 
\end{figure}

Figure 5(a) shows the fluorescence image of a cloud of \mybe ions loaded by applying single pulses of the ablation laser with energies of around \SI{500}{\micro\joule}, using an axial trap frequency $\nu_z\approx\SI{550}{\kilo\hertz}$.
The ablation laser focus diameter on the beryllium target is estimated to be on the order of \SI{60}{\micro\meter}.
Exact determination of the waist diameter is difficult because the laser has a multimode profile and focussing with a parabolic mirror highly depends on beam quality.
Direct creation of ions by ablation from a metallic target has been reported before \cite{chichkov_femtosecond_1996,zimmermann_laser_2012}.
We found that pulse energies of 40 to \SI{80}{\micro\joule} are sufficient to load the trap reliably.
This corresponds to a  peak intensity of about 0.5 to \SI{1}{\giga\watt\per\centi\meter\squared} at the beryllium target's position.
Below a threshold of about \SI{30}{\micro\joule}, no loading is observed.

The loading is insensitive to photoionization laser power (or absence thereof).
The reason for the insensitivity to the photoionization laser is of concern since ions produced directly by the ablation laser are expected to have much larger magnetron radii than ions that are produced in the trap center.
To detect and cool the ions, we shine in laser light with a frequency close to $\nu_0=\SI{957.466784}{\tera\hertz}$ to drive the closed cycling transition $\ket{^2\mathrm{S}_{\nicefrac{1}{2}},\,m_J=\nicefrac{1}{2},\,m_I=\nicefrac{3}{2}}\rightarrow\;\ket{^2\mathrm{P}_{\nicefrac{3}{2}},\,m_J=\nicefrac{3}{2},\,m_I=\nicefrac{3}{2}}$. We observe a strong fluorescence signal on the EMCCD detector after loading the trap.
Figures 5(b) and (c) show images of a large and small cloud of ions after optimization of the imaging system.
When continuously laser-cooled, clouds can be trapped for days.

\subsection{Demonstration of Doppler cooling}

\begin{figure}
	\centering
	\includegraphics[width=\columnwidth]{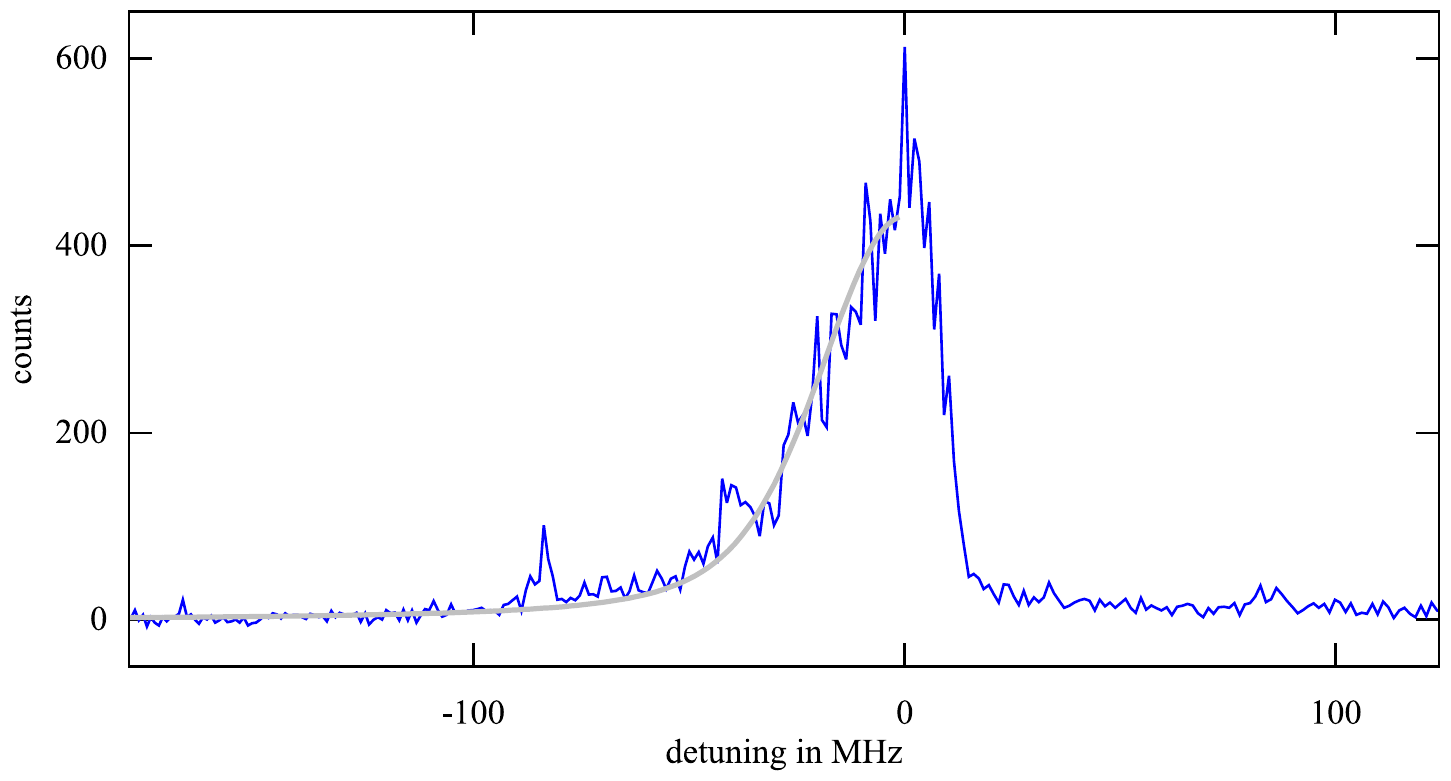}
	\caption[Figure 6]{Background corrected fluorescence signal of a frequency scan with a cloud of ions (blue) and fitted Voigt profile (grey). Scattering processes on the blue-detuned side of the resonance heat the ions, resulting in an asymmetric lineshape.}
	\label{fig8}
\end{figure}

Effective laser cooling in a Penning trap is complicated by the metastable nature of the magnetron motion \cite{itano_laser_1982,thompson_simple_2000}.
The negative energy associated with the magnetron motion makes it hard to cool all three motional modes at the same time.
When referring to `cooling' the magnetron motion, we mean reducing its quantum number, reducing the magnetron radius and hence putting energy into the mode rather than removing it.
In contrast to many other experiments, we use a single laser beam that has an inclination of $45^\circ$ with respect to the trap axis, so it interacts with all motional modes.

Upon ramping the cooling laser to the red-detuned side of the resonance, we observe a distinctive change in the fluorescence spectrum over time.
Directly after loading, the fluorescence is unaffected by the laser frequency over a tuning range of \SI{1}{\giga\hertz} to the red of the resonance.
Assuming Doppler broadening, we estimate a temperature in excess of \SI{1000}{\kelvin}.
After cooling, a clear resonance, as shown in \fref{fig8}, with a width (FWHM) of \SI{47}{\mega\hertz} can be extracted from fitting a Voigt profile \cite{olivero_empirical_1977} to the red-detuned side of the spectrum.
Heating processes lead to a sharp drop of fluorescence on the blue-detuned side of the resonance frequency.

Considering that the natural linewidth of the transition is \SI{19.6}{\mega\hertz} \cite{andersen_mean-life_1969}, we deconvolve the Voigt profile with the corresponding Lorentzian lineshape, and derive a Gaussian FWHM of $\Delta\nu_\text{D}=$\SI{35.5\pm1.0}{\mega\hertz}, which we attribute to Doppler broadening.
The corresponding temperature is \SI{24.0(7)}{\milli\kelvin} using \cite{foot_atomic_2005}
\begin{equation}
T=\frac{(\Delta\nu_\text{D})^2 m c^2}{8\nu_0^2 k_\text{B}\cdot\ln 2}.
\end{equation}
The Doppler limit for the cooling transition used is \SI{0.5}{\milli\kelvin}, hence further optimization of the cooling parameters, such as beam pointing, should allow to further decrease the temperature of the particle.
By moving the laser beam away from the trap center, an intensity gradient over the radial extent of the ion's trajectory is created that allows to cool all three modes simultaneously \cite{itano_laser_1982,thompson_simple_2000}.
We also realized cooling of all three modes by applying a red-detuned laser to cool the axial and cyclotron modes efficiently and simultaneously irradiating a sideband drive with frequency $\nu_\mathrm{z}+\nu_-$ (where $\nu_\mathrm{z}$ is the axial and $\nu_-$ the magnetron frequency) to couple the two modes, thus creating a steady state in which all motional modes are cooled \cite{dehmelt_entropy_1976}.

\subsection{Reduction of particle number}

\begin{figure}
	\begin{center}
		\includegraphics[width=.45\columnwidth]{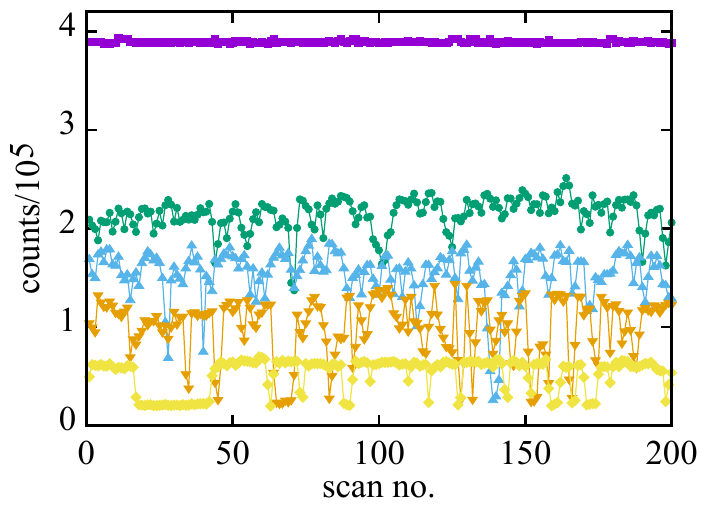}
		\hfill
		\includegraphics[width=.45\columnwidth]{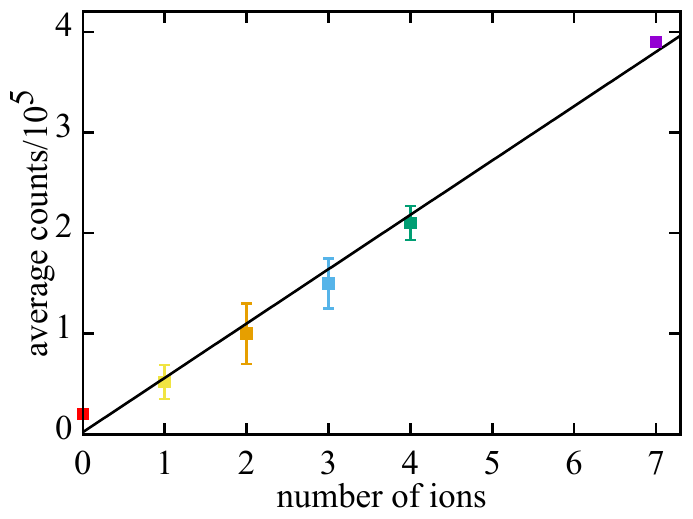}
	\end{center}
	\caption[Figure 7]{(a) Five time-resolved signals of individual clouds using the same trap and laser parameters. Each data point was acquired with an exposure time of \SI{0.5}{\second}. Stepwise decrease of the signal is visible as well as different count levels for each cloud. (b) shows the average of each measurement in (a) plotted over the inferred number of ions with a linear fit to the data points. The red point indicates the measurement background with no ions in the trap.}
	\label{fig7}
\end{figure}

With our current loading scheme, we usually trap a large cloud of ions.
Since we plan on using single \mybe ions for motional coupling and subsequent steps, we need to reduce the number of particles.
Towards that end, we trap a cloud with a high negative voltage on the ring electrode of around \SI{-80}{\volt}.
We then decrease the voltage by a factor of 10 to remove the most energetic particles from the trap.
After some precooling, we ramp the voltage back to the original value to get a tight confinement of the remaining cloud.
We then apply a voltage sequence on the electrodes to adiabatically split the ion cloud and discard a part of it.
The remaining cloud is recooled subsequently.
After a few of these splitting cycles, we end up with a small cloud.
On numerous attempts (see Figure \ref{fig7}), we found a smallest increment in the fluorescence signal which we attribute to a single ion.
Figure \ref{fig7}(b) shows a linear relation between the number of ions and counts on the EMCCD camera of \num{5.4e4} counts per ion using an exposure time of \SI{0.5}{\second} and an overall gain of 78 (electron multipler gain and preamplifier of the camera).

\section{Prospects and discussion}
Now that the trapping infrastructure and several essential steps such as loading, cooling, and detection have been implemented successfully, we pursue the goal of demonstrating motional coupling of two individual ions in separate potential wells in a Penning trap for the first time.
Since it is crucial for the coupling scheme that the axial trap frequencies are equal, our first goal is to demonstrate coupling between two \mybe ions in a symmetric potential.
In addition, we are currently testing a proton source trap that will allow proton loading without blocking on-axis access to the trap.
Moreover, we are setting up the associated electronics for work with protons, such as detection tank circuits and amplifiers \cite{nagahama_highly_2016}.\\
Furthermore, we are working towards sub-Doppler cooling of \mybe in a Penning trap.
While sideband cooling in a Penning trap has been demonstrated \cite{goodwin_resolved-sideband_2016}, the used $^{40}\mathrm{Ca}^+$ ions are amenable to direct sideband cooling using a single laser, as an optical qubit transition is used. For \mybe, a Raman process is needed to interact with the motional modes on the single quantum level. Towards that end, the ground state hyperfine splitting must be bridged using two phase-coherent lasers with a frequency difference of around \SI{140}{\giga\hertz}.
This frequency gap is too large to be closed using acousto-optic modulators, as it is common to do in low-field applications.
We have recently demonstrated the use of a pulsed laser \cite{hayes_entanglement_2010} to drive motional sideband transitions in \mybe \cite{paschke_versatile_2019}, an approach which can bridge this big qubit splitting. 

On longer timescales, miniaturisation of the coupling trap using established techniques of microfabrication is pursued in order to increase the coupling rate.
Together with ground-state cooling, manipulation of the motional states is in reach as well as quantum logic spectroscopy, potentially shortening the preparation time of single antiprotons with spin state detection fidelity of $>\SI{98}{\percent}$ by more than two orders of magnitude \cite{smorra_base_2015}.

Precision experiments in Penning traps usually utilize the Brown-Gabrielse invariance theorem \cite{brown_precision_1982} to determine the free cyclotron frequency $\nu_c$ of the particle of interest.
In precision studies on the p.p.b$.$-level and below, energy dependent systematic shifts of the eigen-frequencies of the three Penning trap modes need to be characterized and corrected.
These shifts arise from trap imperfections due to non-linear terms in the trapping potential, inhomogeneities of the magnetic field, image charge and relativistic shifts, and have been summarized in \cite{ketter_first-order_2014,brown_geonium_1986}.
Of leading concern are shifts of the modified cyclotron frequency $\nu_+(E_+,E_-,E_z)$.
Here, $\nu_+$-shifts, imposed by the weakly bound axial oscillator, constitute a dominant source of systematic uncertainty.
The shift imposed by a non-vanishing magnetic bottle term $B_2$ of a slightly inhomogeneous magnetic field is given by
\begin{eqnarray}
\frac{\Delta\nu_+}{\nu_+}= \frac{1}{4\pi^2m \nu_z^2}\cdot\frac{B_2}{B_0}\cdot E_z.
\end{eqnarray}
For protons at a typical axial frequency $\nu_z=\SI{650}{\kilo\hertz}$ in a magnetic field of $B_0=\SI{2}{\tesla}$ \cite{smorra_parts-per-billion_2017,schneider_double-trap_2017}, the fractional systematic $\nu_+$-shift is 
\begin{equation}
\frac{\Delta\nu_+}{\nu_+}= 2.5\cdot10^{-10}\cdot B_2\cdot T_z,
\end{equation}
where $T_z$ is the axial temperature of the particle.
For example in non-compensated g-factor experiments $B_2$ is typically in a range between \SI[per-mode=symbol]{0.2}{\tesla\per\meter\squared} and \SI[per-mode=symbol]{4}{\tesla\per\meter\squared}, while $T_z\approx\SI{10}{\kelvin}$ \cite{schneider_double-trap_2017,smorra_parts-per-billion_2017}.
Depending on the chosen frequency measurement method, the  $\Delta\nu_+(B_2,T_z)$-scaling can induce a dominant systematic error \cite{smorra_parts-per-billion_2017}.
With the axial temperature achieved in this work, and by imprinting it to the axial mode of a co-trapped antiproton in advance to a cyclotron frequency measurement, the related cyclotron frequency shift would be suppressed to a level of $<20\,$ppt.

The axial energy exchange between an antiproton and a laser cooled Be-ion is expected to take about \SI{60}{\milli\second}.
Using a sideband pulse \cite{cornell_mode_1990}, the axial energy can be translated to cyclotron mode energy, achieving $T_+=(\nu_+/\nu_z)T_z\approx \SI{2}{\kelvin}$ with the trap parameters currently used.
Using this scheme, a single antiproton at $T_z<\SI{0.1}{\kelvin}$, and therefore a particle with a spin-state identification fidelity $>\SI{95}{\percent}$ \cite{smorra_observation_2017}, could be prepared in an effective coupling time of about \SI{2}{\second}.
Compared to recent BASE antiproton magnetic moment measurements, this would reduce the effective $\nu_+$-mode temperature by about a factor of 5, and the coupling time to thermalize the antiproton by more than a factor of 100.
This advance would eliminate the dominant time-consuming step of the measurement sequence, will enable measurements at significantly improved sampling rate and substantially lower systematic error \cite{smorra_parts-per-billion_2017,schneider_double-trap_2017}. \\
Note that the temperature which was achieved for the beryllium cloud above is an ensemble temperature with broadened recoil parameter and therefore increased temperature.
We expect that with a single particle, the axial temperature can be reduced to the Doppler limit, a factor of 40 lower \cite{sawyer_spin_2014}.
This will considerably improve the performance of future magnetic moment experiments.

\section{Conclusion}
We have designed and commissioned a novel cryogenic Penning trap system for \mybe ions.
We have successfully demonstrated ablation loading, Doppler cooling, and reduction of the number of ions down to a single particle which has been detected unambiguously.
We have outlined the next steps towards motional coupling and sympathetic cooling of two ions that can be generalized to arbitrary charged ions.
We anticipate a high impact of the sympathetic cooling scheme on precision measurements in Penning traps, such as $q/m$- or $g$-factor measurements on antiprotons and protons, molecular ions or other systems that are not amenable for direct laser cooling.

\ack
We would like to thank the Laboratory of Nano and Quantum Engineering (LNQE) of Leibniz Universit\"at Hannover, the IQ machine shop and the PTB workshop for support.
We acknowledge funding by the ERC through StG `QLEDS' and DFG through SFB 1227 `DQ-mat' and RIKEN - Pioneering Project Funding.

\section*{References}
\bibliography{ta-12}

\end{document}